\documentclass[10pt, conference,twocolumn]{ieeeconf}                                                      

\IEEEoverridecommandlockouts                                                                               
\usepackage{graphics} 
\usepackage{epsfig} 
\usepackage{cite}	
\usepackage{dsfont}
\usepackage{amsmath, amssymb} 
  \usepackage{amsthm}
\usepackage{makecell}
\usepackage[dvipsnames]{xcolor}
\usepackage[overload]{empheq}
\usepackage[hyperindex=true, %
  pdftitle={Comparability in Generalized Nash Equilibrium Problems},%
  pdfauthor={Authors},%
  colorlinks=true,%
  pagebackref=false,%
  plainpages=false,%
  pdfpagelabels,%
  linkcolor=black,%
  citecolor=black,%
  filecolor=black,%
  urlcolor=black%
  ]{hyperref}

\usepackage{booktabs}
\usepackage{tablefootnote}
\usepackage{subcaption}
\usepackage[normalem]{ulem}

\definecolor{lightgray}{gray}{0.9}
\definecolor{ggreen}{rgb}{0,0.5,0}
\definecolor{rred}{rgb}{0.5,0,0} 

\newcommand{\mc}[1]{\mathcal{#1}}

\newcommand{\R}{\mathbb{R}}

\DeclareMathOperator*{\argmin}{arg\,min}

\DeclareMathOperator{\vcol}{col}
\DeclareMathOperator{\subjectto}{\ {s.}{t.}\ }

\newtheorem{ass}{Assumption}

\newtheorem{dfn}{Definition}

\newtheorem{prp}{Proposition}
\newtheorem{cor}{Corollary}

\newcommand{\sh}[1]{\textcolor{NavyBlue}{[#1]\raise 0.5ex \hbox{\footnotesize{SH}}}}
\newcommand{\hn}[1]{\textcolor{Orange}{[#1]\raise 0.5ex \hbox{\footnotesize{HN}}}}
\newcommand{\sav}[1]{\textcolor{Green}{[#1]\raise 0.5ex \hbox{\footnotesize{SB}}}}

\graphicspath{{figs/}}

\title{\LARGE \bf
The Limits of ``Fairness'' of the \\
Variational Generalized Nash Equilibrium}

\author{Sophie Hall, Florian D\"orfler, Heinrich H. Nax, Saverio Bolognani
\thanks{The authors are with the ETH Zurich Automatic Control Lab, 8092 Zurich, Switzerland and University of Zurich, Behavioral Game Theory, 8050 Zurich, Switzerland. {Emails: \texttt{\{shall, dorfler, hnax, bsaverio\}@ethz.ch}}. This work was supported by the Swiss National Science Foundation under the NCCR Automation (grant 51NF40\textunderscore225155) and under the Eccellenza Grant "Markets and Norms" (HN).}
}

\begin{document}

\maketitle
\thispagestyle{empty}
\pagestyle{empty}

\begin{abstract}

Generalized Nash equilibrium (GNE) problems are commonly used to model strategic interactions between self-interested agents who are coupled in cost and constraints. Specifically, the variational GNE, a refinement of the GNE, is often selected as the solution concept due to its non-discriminatory treatment of agents by charging a uniform ``shadow price" for shared resources.
We study the fairness concept of v-GNEs from a comparability perspective and show that it makes an implicit assumption of unit comparability of agent's cost functions, one of the strongest comparability notions. Further, we introduce a new solution concept, f-GNE in which a fairness metric is chosen \textit{a priori} which is compatible with the comparability at hand. We introduce an electric vehicle charging game to demonstrate the fragility of v-GNE fairness and compare it to the f-GNE under various fairness metrics.

\end{abstract}

\section{Introduction} 


Access to resources, including natural resources (water, air, etc.) and infrastructures (energy, traffic, communication, health services, etc.), is increasingly automatically controlled by algorithms, often distributed through large-scale systems. 
We can think of the individuals seeking access to these resources as self-interested agents, each wanting to fulfill their needs at minimal cost. In control-theoretical works, such settings are often modeled as generalized Nash equilibrium (GNE) problems~\cite{belgioioso2022distributed, ma2016efficient,benenati2024probabilistic}. GNEs are different from classical non-cooperative Nash games, where coupling occurs only in the cost functions, because they also allow for coupling in the agents' constraints, which is the important feature that allows to model access to a shared but finite resource~\cite{facchinei2007gnep}. 

The most common solution concept used for generalized games is the variational GNE (v-GNE), which is characterized as the solution to a variational inequality formed from the coupling constraint and the pseudo-gradient mapping of agents' cost~\cite{facchinei2007gnep}. Specifically, v-GNE constitutes a refinement of GNE that requires that the Lagrange multiplier associated with the coupling constraint is uniform across agents.

V-GNEs have been used in many applications, see Table~\ref{tab:vGNEsLit}. They are useful for real-time control by algorithms in multi-agent settings due to the following desirable properties. 

\begin{figure}[t]
\centering
\includegraphics[width=0.85\columnwidth]{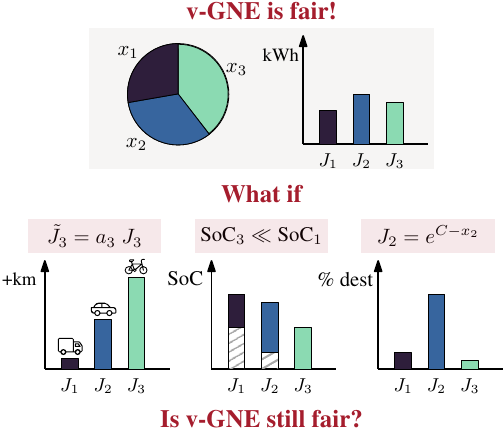}
\caption{Pictorial representation of the effect of the high level of interpersonal comparability assumed by the v-GNE solution concept. 
Additional information (that can be unmeasurable, misreported, or deemed extraneous) affects the fairness of the resulting allocation (cf. Section~\ref{ssec:EVtransformations}).}
\label{fig:fairness}
\end{figure}

Firstly, as they are the solution to a variational inequality, existence and uniqueness results are readily available under a broad set of assumptions~\cite{rosen1965existence, facchinei2007gnep}. This fact also permits the deployment of fast algorithms solving for v-GNEs based on different semi-decentralized \cite{paccagnan2019nash, belgioioso2023semi} or fully-distributed \cite{yi2019operator, bianchi2022fast} communication structures. 
While there exist algorithms to solve for other solutions within the set of GNEs than v-GNEs, the initialization of such algorithms has a direct influence on the converged GNE solution~\cite{kaisermayer2021operation, facchinei2009generalized}, which constitutes a drawback if we want to characterize properties of the solution \emph{ex ante}.
Secondly, as v-GNEs are GNEs, they are inherently strategically stable. That is, they constitute outcomes at which no agent has an incentive to deviate unilaterally.
Thirdly, prior works have established explicit bounds and guarantees on the  efficiency of v-GNEs~\cite{kulkarni2019efficiency,cadre2020peer} in terms of utilitarian sum of costs, while GNEs may be arbitrarily inefficient.
%



Lastly, but arguably most importantly according to the v-GNE literature \cite{dreves2018generalized, benenati2024probabilistic, bassanini2002allocation}, the main appeal for why v-GNEs constitute the most desirable solutions within the set of GNEs is their fairness. 
Fairness in v-GNEs boils down to an equal penalty, or common `shadow price' to give it an economic interpretation, that is associated with access to the resource in terms of the associated uniform Lagrange multipliers. By contrast, at a general GNE solution, these Lagrange multipliers (i.e. shadow prices or marginal costs) may vary across agents, which is deemed ``unfair'' in settings in which individuals are anonymous or indistinguishable to the planner~\cite{kulkarni2012variational, harker1991generalized}. 

Fairness interpreted more broadly is a concept often assessed in terms of the trade-off between efficiency and equity of a solution (e.g., via Atkinson's~\cite{atkinson1970measurement, bertsimas2012efficiency}, Theil's~\cite{theil1967economics} or Jain's index~\cite{jain1984quantitative}). Indeed, by now many fairness metrics have been used in control~\cite{lan2010axiomatic, cadre2022parametrized,villa2025fair,neely2008fairness} that are suitable for optimization-based techniques~\cite{xinyingchen2023guide} beyond these classical indices such as maximin, Gini coefficient, alpha fairness, etc.~\cite{xinyingchen2023guide}.
Despite the frequent motivation of v-GNEs based on a fairness argument, we are not aware of prior research investigating how equal shadow prices of v-GNEs relate to more general fairness notions, and under which assumptions they are fair. 
Our work is inspired by a recent numerical study of fair equilibrium selection from within the set of GNEs as conducted by~\cite{cadre2022parametrized} that focused on Jain's fairness index. 
This work pointed to a gaping research lacuna, as we still know virtually nothing about the real fairness properties of the v-GNE in general, requiring principled theory and more numerical studies. We do not aim to fill this gap here but rather make explicit under which set of fundamental assumptions the ``fairness'' paradigm of equal pricing in v-GNEs is meaningful and when it is necessary to be cautious.



We approach this challenge from a comparability angle, a notion studied in formal social choice theory~\cite{roberts1980interpersonal, daspremont2002social}. 
Comparability defines to what extent the costs incurred by different agents are commensurable.
By deciding a level of comparability, a planner decides what information about the individual costs is relevant to the resource allocation problem. 
Conversely, a limited level of comparability can be seen as an invariance condition: it dictates what information about the individual costs (e.g., scale, offset) should not affect the resource allocation. See~\cite{shilov2026welfare} for a general framework for comparability in control theory.
Adequately modeling comparability is of course particularly crucial when optimizing for metrics such as fairness, equality, and equity: because only by defining \textit{a priori} what can be considered ``equal'' can we optimize for any such metric meaningfully.






We prove that the solution set of GNEs does not rely on cardinal comparability of the agents' costs because of their very nature of being competitive equilibria. However, v-GNEs require a restrictive notion of comparability, as they fundamentally and inevitably tradeoff costs: they are a coordinated decision, not the result of agent's autonomous decisions. 
The fairness of the resulting tradeoff depends on whether the agents' costs are unit-comparable.
This difference between GNEs and v-GNES has been pointed out in \cite{facchinei2007gnep} and \cite{dreves2018how} as a matter of scale invariance, but it has not been connected to comparability or fairness.

To summarize, the contributions in this paper are:

\begin{itemize}
\item[(i)] To the best of our knowledge, we are the first to apply comparability notions to solution concepts outside of social choice theory, specifically to GNEs. 
\item[(ii)] We demonstrate that selecting the v-GNE assumes at least cardinal unit comparability of agents' cost, which is one of the strongest comparability notions.
\item[(iii)] We propose a new solution concept called the f-GNE, which maximizesan \textit{a-priori} defined fairness metric.
\item[(iv)] We highlight the fragility  of v-GNEs' fairness and
compare v-GNEs to other fairness metrics in an electric
vehicle charging game.
\end{itemize}


\begin{table}[t]
\centering
\def\arraystretch{1.1}
\begin{tabular}{@{}lll@{}}
\textbf{Application} & 
\textbf{Justification} & 
\textbf{Publication}    \\ 
    \toprule
Utility-based network optim.  & Comput. eff., Str. stable  & \cite{kelly1998rate,pan2009games}    \\
Autonomous driving  & Fair &  			\cite{dreves2018generalized, lecleach2022algames}   \\ 
Railroad track-time pricing  &  Comput. eff., Fair  &          \cite{bassanini2002allocation}                  \\
Energy management &  Comput. eff., Fair &\cite{belgioioso2022operationally, behrunani2023designing, jingyuan1999spatial, cadre2020peer, hall2022receding}            \\
Supply chains    &  Fair &     \cite{hall2024game, belgioioso2025online}                       \\
Wireless communication & - &  \cite{yin2011nash}                           \\
Environmental games   & -  &     \cite{krawczyk2005coupled, bahn2008class}        \\             
Electric vehicle charging & Existence \& uniq.  & \cite{ma2016efficient, ma2013decentralized}\\
Traffic routing & Fair & \cite{benenati2024probabilistic} \\
\bottomrule\end{tabular}
\caption{Control applications employing v-GNEs.}
\label{tab:vGNEsLit}
\end{table}
%
%


\section{Preliminaries} 
\label{sec:preliminaries}



\subsection{Generalized Nash equilibrium problems}

Consider a set of self-interested agents $i\in\mc{I}:=\{1,\dots,M\}$ each minimizing their cost function $J_i(x_i,x_{-i})$, which depends on their own decision $x_i \in \R^{n_i}$ and other agents' decisions $x_{-i}$. 
We denote by $x$ the vector obtained by stacking all agents' decisions, i.e., $x = \vcol(\{x_i\}_{i\in \mc{I}})\in \R^n$.
We will make the following assumption for the entire paper.

\begin{ass}[Convex differentiable cost functions]
The cost functions $J_i(x_i,x_{-i})$ are strongly convex and continuously-differentiable in $x_i$, for every fixed $x_{-i}$.
\label{ass:convexJi}
\end{ass}
\noindent Further, each agent's feasible set 
\begin{align}
\mc{C}(x_{-i}):= \{x_i \in \R^{n_i}~|~ g_i(x_i, x_{-i})\leq 0\}
\label{eq:GNEconstraints}
\end{align} 
depends on the decision of other agents with $g_i(\cdot): \R^n \to \R^m$, yielding a set of coupled optimization problems:
\begin{equation} \label{eq:GNEP1}
\forall i\in \mc{I}: \quad \min_{x_i}\, {J_i(x_i, x_{-i})} \subjectto x_i \in \mc{C}_i(x_{-i}). 
\end{equation}
%

\begin{ass}[Convex constraint sets] \label{ass:constraintSets}
The sets $\mc{C}_i(x_{-i})$ are closed and convex and the constraint functions $g_i(\cdot, x_{-i})$ are continuously differentiable. 
\end{ass}

Decisions that jointly solve~\eqref{eq:GNEP1} are generalized Nash equilibria (GNE)~\cite[\S 2]{facchinei2009nash}, as defined next.
\begin{dfn}[Generalized Nash Equilibrium] \label{dfn:GNE} A joint decision $x^*$ is a GNE  of~\eqref{eq:GNEP1} if $\,\forall i \in \mc{I}:$
\begin{align*}
J_i(x_i^*, x_{-i}^*) \leq J_i(x_i, x^*_{-i}), \quad \forall x_i \in \mc{C}_i(x_{-i}^*),
\end{align*} with the solution set denoted as $\mc{S}^{\text{\tiny GNE}}$.
\end{dfn}
%
%
Intuitively, a GNE is a joint decision for which no agent $i \in \mc{I}$ can reduce its cost by unilaterally changing its own decision. Under Assumptions~\ref{ass:convexJi} and \ref{ass:constraintSets}, finding a GNE is equivalent to solving the quasi-variational inequality
\begin{align}\label{eq:QVI}
(x-x^*)^\top F(x^*)\geq 0,  \quad \forall \,x  \in \mc{C}(x^*)
\end{align}
where $F(x) = \vcol( \{\nabla_{x_i} J_i(x_i, x_{-i})\}_{i\in \mc{I}})$ and $\mc{C}(x) := \prod_{i\in\mc{I}} \mc{C}_i(x_{-i})$. 
Equivalently, $x^*$ is a GNE if a suitable constraint qualification holds (e.g., the Mangasarian-Fromovitz or the Slater's condition) and if there exist multipliers $\{\lambda_i \in \R^{m_i} \}_{i\in\mc{I}}$ such that the following KKT conditions hold
\begin{align}\label{eq:KKTGNE}
\forall i\in \mc{I}:\; \left\{
\begin{array}{ll}
\nabla_{x_i}J_i(x^{*}) + \nabla_{x_i}g_i(x^*)^\top \lambda_i^*=0\\
0\leq \lambda_i^* \perp -g_i(x^*)\geq 0 .
\end{array}
\right. 
\end{align}
In control applications, the majority of works focuses on setups where the coupling constraints are \textit{jointly convex}~\cite[Def. 3.6]{facchinei2007gnep}, i.e., the following additional assumption holds.

\begin{ass}[Jointly convex constraints]
There exists a non-empty convex set $C$ such that 
$ \mc{C}_i(x_{-i}):= \{x_i ~|~ (x_i,x_{-i})\in C \}$.    
\label{ass:jointlyconvexconstraints}
\end{ass}

Notice, in this case, the $g_i(x_i,x_{-i})$ defining $\mc{C}(x_{-i})$ is the same across agents, $g_i = g, \forall i$.
Under Assumptions~\ref{ass:convexJi} and \ref{ass:jointlyconvexconstraints}, the subclass of \emph{variational} GNEs (v-GNEs) can be defined.

\begin{dfn}[Variational GNE] \label{dfn:vGNE} 
Consider the GNE problem~\eqref{eq:GNEP1}, and let Assumptions~\ref{ass:convexJi} and \ref{ass:jointlyconvexconstraints} hold.
The joint decision $x^*$ is a v-GNE if
\begin{align}\label{eq:vGNEsol}
(x - x^{*})^\top F(x^*)\geq 0  \quad \forall x  \in C
\end{align}
and the solution set is denoted by $\mc{S}^{\text{\tiny v-GNE}}$.
\end{dfn}
Equivalently, under appropriate constraint qualifications and assuming continuous differentiability of the function $g$, $x^*$ is a v-GNE if there exists a multiplier $\lambda \in \R^{m}$ such that the following KKT system is satisfied
\begin{align}\label{eq:KKToriginalvGNE}
\forall i\in \mc{I}:\; \left\{
\begin{array}{ll}
\nabla_{x_i}J_i(x^{*}) + \nabla_{x_i}g(x^*)^\top \lambda^{*}=0\\
0\leq \lambda^* \perp -g(x^*)\geq 0 
\end{array}
\right. 
\end{align}
Note the uniform Lagrange multiplier and the uniform constraint function $g$ across agents. In general, even in the case of jointly convex constraints, the set of GNE solutions $\mc{S}^{\text{\tiny GNE}}$ is large, potentially infinite. The v-GNE is a refinement of the GNE, that is, $ \mc{S}^{\text{\tiny v-GNE}}\subseteq  \mc{S}^{\text{\tiny GNE}}$~\cite{kulkarni2012variational,fukushima2011restricted}.

\subsection{Inter-Agent Cost Comparability}

To determine the fairness properties of a joint decision $x$, it is necessary to compare the cost incurred by different individual agents.
In different contexts, it may be that it is meaningful only to compare the absolute costs, to compare the relative improvements, or to compare costs with respect to a worst case scenario. What is comparable in a specific application depends on how difficult it is to measure certain information, on whether self-reporting can be trusted, and on what information the social planner considers relevant (see~\cite{shilov2026welfare} for further discussion in the context of control). 

This notion of comparability can be expressed by defining a concept of equivalence between cost functions, i.e., determining what cost functions cannot or should not be distinguished.
Equivalently, a specific level of comparability corresponds to a family of transformations $\phi(\cdot)$ of the cost functions that must not affect the joint decision (i.e., for which the decision must be invariant).

\begin{figure}[tb]
\centering
\includegraphics[width=0.45\columnwidth]{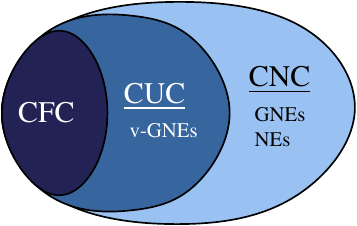}
\caption{Solution concepts for competitive games without (NE) and with coupling constraints (GNE, v-GNE) - each with their respective comparability assumptions.}
\label{fig:vennGNE}
\end{figure}

In the following, we review a selection of comparability levels commonly used in social choice theory \cite{roberts1980interpersonal}, we report the corresponding invariance transformation, and we provide a brief interpretation. Figure~\ref{fig:vennGNE} illustrates the hierarchy of these comparability levels.

\medskip

\noindent\textbf{Cardinal Non-Comparability (CNC)}:
 \begin{align}\label{eq:CNC}
\phi_{\text{\tiny CNC}}(J_i(x)) :=  a_i\,J_i(x) \;+\; b_i,
\quad a_i>0,\; b_i\in\mathbb{R}.
 \end{align}
%
Hence, we cannot compare costs or cost increments across agents, as the coefficients $a_i$ and $b_i$ are heterogeneous.

\medskip

\noindent\textbf{Cardinal Unit Comparability (CUC)}:
 \begin{align}\label{eq:CUC}
\phi_{\text{\tiny CUC}} (J_i(x)) :=  a\,J_i(x) \;+\; b_i, \quad a>0,\; b_i\in\mathbb{R}.
 \end{align}
As $a$ is uniform, we can compare cost increments across agents. However, the absolute cost levels are not comparable. 

\medskip

\noindent\textbf{Cardinal Full Comparability (CFC)}:
\begin{equation}
 \phi_{\text{\tiny CFC}} (J_i(x)) := a\,J_i(x) \;+\; b, \quad a>0,\; b\in\mathbb{R}.
\end{equation}
A single absolute scale is shared by all agents, so both cost levels and cost increments are comparable.

\smallskip

\section{Comparability in Generalized Nash Equilibrium Problems}
\label{sec:GNEcomparability}

Consider the generalized game~\eqref{eq:GNEP1} with coupling constraint sets~\eqref{eq:GNEconstraints} and the problem of finding $x^{*}\in \mc{S}^{\text{\tiny GNE}}$, as stated in Definition~\ref{dfn:GNE}. Let Assumptions~\ref{ass:convexJi} and \ref{ass:constraintSets} hold true. Further, assume that for every agent $i$, a suitable constraint qualification holds at $x^*$. The following result relates GNE solutions to the cardinal comparability of agents' costs.


%
%

%
%

\begin{prp}[Invariance of GNE under CNC]\label{prp:CNCofGNE}
Let Assumptions~\ref{ass:convexJi} and \ref{ass:constraintSets} hold.
The solution set $\mc{S}^{\text{\tiny GNE}}$ of the GNE problem~\eqref{eq:GNEP1} is invariant under any transformation $\phi_{\text{\tiny CNC}}(\cdot)$ defined in~\eqref{eq:CNC}.
\end{prp}

\begin{proof}
The proof is given in Appendix~\ref{sec:App_Prop1}.
\end{proof}


\begin{cor}[Invariance of NE under CNC]
The solution set $\mc{S}^{\text{NE}}$ of a Nash equilibrium problem without coupling constraints
is invariant under transformations $\phi_{\text{\tiny CNC}}(\cdot)$ in \eqref{eq:CNC}.
\end{cor}
\begin{proof}
This is a direct application of Proposition~\ref{prp:CNCofGNE} when the sets $\mc{C}_i(x_{-i})$ are independent of $x_{-i}$.
\end{proof}


Proposition~\ref{prp:CNCofGNE} shows that the solution concept of GNE requires no comparability between the costs incurred by different agents.
This is not the case when we consider the refinement to v-GNEs, as the following results show.

\begin{prp}[Invariance of v-GNE under CUC]\label{prp:CUCofVGNE}
Let Assumptions~\ref{ass:convexJi}-\ref{ass:jointlyconvexconstraints} hold.
The solution set $\mc{S}^{\text{\tiny v-GNE}}$ of the GNE problem~\eqref{eq:GNEP1} is generally not invariant under transformations $\phi_{\text{\tiny CNC}} (\cdot)$ defined in~\eqref{eq:CNC}. $\mc{S}^{\text{\tiny v-GNE}}$ is invariant under $\phi_{\text{\tiny CUC}} (\cdot)$ defined in~\eqref{eq:CUC}.
\end{prp}
\begin{proof} The proof is given in Appendix~\ref{sec:App_Prop2}.
\end{proof}



The fact that v-GNE, as a solution concept, requires at least Cardinal Unit Comparability (see Figure~\ref{fig:vennGNE}) is particularly important when fairness considerations are invoked to justify it as a refinement of the larger set of GNE solutions.
When doing so, the social planner needs to ensure that the comparability of cost increments is possible and meaningful.
For example, they need to ensure that (i) the elicitation of cost functions from individuals is robust against strategic manipulation by the agents;  (ii) the class of admissible cost functions is rich enough to represent the real cost (more precisely: cost increments) incurred by the agents;
and (iii) that enough information from the agents is collected to calibrate the relative scaling of each agent's cost function.


\begin{table*}[tb] 
    \centering
    \renewcommand{\arraystretch}{1.8}
    \setcellgapes{3pt}
    \makegapedcells
    \begin{tabular}{@{}ccccc@{}}
        \textbf{Fairness metric}  & \textbf{Function} $f(\cdot)$& \textbf{Interpretation}  & \textbf{Comparability}\\
    \toprule
        Rawlsian Maximin (MM)  \cite{rawls1971theory}& $\displaystyle \max_{i\in \mc{I}} J_i(x)$ &  Optimizing the outcome for the worst-off. & CFC \\ 
        Social welfare or utilitarianism (SW) &   $\displaystyle \sum_{i\in \mc{I}} J_i(x)$  &\makecell{Focuses on efficiency, an allocation that minimizes\\ total costs, regardless of how it is distributed.} & CUC \\ 
        Nash Bargaining solution (NBS)~\cite{nash1950bargaining} & \makecell{$ \displaystyle -\prod_{i\in \mc{I}} \left[J_i (\hat x) - J_i(x)) \right]$} & \makecell{Promotes equity in terms of proportional cost\\improvements compared to the benchmark.} & CNC\\
        \makecell{Alpha fairness or\\ Atkinson's index (AI)~\cite{atkinson1970measurement, bertsimas2012efficiency}}& 
$\displaystyle \sum_{i\in \mc{I}} \frac{\left[ J_i(x) \right]^{1 - \alpha}}{1 - \alpha}, \  \substack{\alpha \neq 1\\\alpha>0}$
& \makecell{Allows to trade-off efficiency \& equity through $\alpha$\\
 $\alpha\to \infty$: (MM), $\alpha = 0$: (SW), $\alpha = 1$: (NBS) } &  CFC \\  
        Jain's index (JI)~\cite{jain1984quantitative} & $\frac{\big( \sum_{i\in \mc{I}} ( J_i(x)\big)^2}{M \sum_{i\in \mc{I}} (J_i(x))^2}$ &  \makecell{Minimizes relative variations across agents.\\ Specifically developed for telecommunications.} & CFC\\
        \bottomrule
     
    \end{tabular}
    \caption{Fairness notions (trading-off equity and efficiency) used in the control literature and their assumed comparability. $J_i (\hat x) $ are the costs of a benchmark outcome $\hat x$ and can be set to zero if appropriate, i.e., if $J_i (\hat x)=0 \ \forall i$.\label{tab:FairnessMetrics}}
    \label{table}
\end{table*}

\section{Fair Generalized Nash Equilibrium (f-GNE)}
\label{sec:fgne}


As discussed in the Introduction, the set of GNEs is large and selecting one is necessary in real implementations. A common goal is to design a solution concept which protects individual self-interest, ensures strategic stability (thus, a GNE), and \textit{additionally} optimizes a desired fairness metric consistent with the assumed inter-agent cost comparability. 
As pointed out previously, v-GNE, as a GNE-refinement, does not generally achieve that. Instead, in line with this observation and inspired by the work in~\cite{cadre2022parametrized}, we introduce a new approach for ``fair'' equilibrium selection that consists in solving the bilevel optimization problem
\begin{subequations}\label{eqn:fGNE}
\begin{align}
x^{\text{\tiny f-GNE}} \in \argmin_{x^*}&\quad f(x^*)\\
\subjectto & \quad x^* \in  \mc{S}^{\text{\tiny GNE}}\label{eqn:fGNE_GNEset}
\end{align}  
\end{subequations}
where $f$ is a fairness metric that, when minimized, defines the desired efficiency-equity tradeoff. Selection of GNEs has been studied in~\cite{benenati2023optimal} from an algorithmic perspective, in contrast to the fairness perspective considered here

In Table~\ref{tab:FairnessMetrics} we summarize fairness metrics commonly used in resource allocation problems in control applications and give a short explanation in which settings they may be appropriate. Interestingly, the v-GNE is a solution of~\eqref{eqn:fGNE} under a specific problem structure and choice of $f$.
\begin{prp}\label{prp:vGNEequalfGNE}
Let Assumptions~\ref{ass:convexJi}-\ref{ass:jointlyconvexconstraints} hold and assume the cost functions of agents are fully decoupled, i.e., $J_i(x)$ is $J_i(x_i)$, and utilitarian fairness is considered with $f=\sum_{i\in \mc{I}} J_i(x)$, then $x^{\text{\tiny v-GNE}} = x^{\text{\tiny f-GNE}}$.
\end{prp}
\begin{proof}
The proof is given in Appendix~\ref{sec:App_Prop3}.
\end{proof}
Note that the connection between the v-GNE solution and the social welfare optimum under decoupled costs has been previously pointed out in~\cite{cadre2020peer} and~\cite{moret2021loss}.

The main difficulty with~\eqref{eqn:fGNE} is that the constraint~\eqref{eqn:fGNE_GNEset} corresponds to the KKT system~\eqref{eq:KKTGNE}. When there are multiple coupling constraints, there exists no closed form expressions of $\mc{S}^{\text{\tiny GNE}}$ and even approximating it numerically is a challenge~\cite{nabetani2011parametrized}. 
%
%
%
However, if there is only one coupling constraint which is separable, the set can be characterized via normalized equilibria. 

\begin{dfn}[Normalized equilibrium] \cite[Dfn. 3.2]{nabetani2011parametrized} 
A GNE $x^*$ is called a normalized equilibrium if there exist $\forall i \in \mc{I}: (x^*, \lambda^*_i)$ which solve~\eqref{eq:KKTGNE} and $ \{r_i>0\}_{i \in \mc{I}}$  satisfying
\begin{align*}
r_1 \lambda_1^*= r_2 \lambda^*_2 = \dots = r_M \lambda^*_M.    
\end{align*}
We say $x^*$ is associated with $r = \{r_i\}_{i \in \mc{I}}\in \R^{m}_{>0}$.
\end{dfn}
 The v-GNE is a normalized equilibrium with $r = \mathds{1}_{m}.$

\begin{prp}\cite[Prp. 3.3]{nabetani2011parametrized}
Let Assumptions~\ref{ass:convexJi}-\ref{ass:jointlyconvexconstraints} hold and suppose $g(\cdot) = \sum_{i\in\mc{I}} g_i(\cdot) : \R^{n}\to \R$ . If $x^*$ is a GNE satisfying strict complementarity with some $\{\lambda_i^*\}_{i \in \mathcal I}$ (i.e., $g_i (x^*) = 0$ implies $\lambda_i^* >0, \; \forall i\in \mc{I}$) then $x^*$ is a normalized equilibrium.
\end{prp}

Consequently, in the case of a single shared constraint, nearly the whole set $\mc{S}^{\text{\tiny GNE}}$ can be characterized by solving 
\begin{align}\label{eqn:VI-NNE}
(x - x^{*})^\top F_r(x^*)\geq 0  \quad \forall x: ~ g(x)  \leq 0 
\end{align}
where $F_r(x)= \vcol( \{\nabla_{x_i} r_i J_i(x_i, x_{-i})\}_{i\in \mc{I}})$ is parametrized by the weights\footnote{Existence and uniqueness results of a solution of~\eqref{eqn:VI-NNE} usually require (strong) monotonicity of $F_r(x)$ which may not hold $ \forall r\in \R^{m}_{>0}$.} $r$. This allows to simplify the constraint~\eqref{eqn:fGNE_GNEset} as presented in the following example. 

\subsection*{An example of f-GNE}


In order to reduce the informational requirements for the f-GNE solution, we choose a fairness metric that is applicable in a setting of cardinal noncomparability, specifically the Nash Bargaining solution (NBS). Using~\eqref{eqn:fGNE} with the NBS metric ensures that 
$x^{\text{\tiny f-GNE}} $ is invariant under CNC transformations defined in~\eqref{eq:CNC}. If we only have one coupling constraint, we can write~\eqref{eqn:fGNE} as follows:
\begin{subequations}
\begin{align}
x^{\text{\tiny f-GNE}} \in \; &\argmin_{r, x^*} - \prod_{i\in \mc{I}} (J_i(\hat{x}) - J_i(x^*))\\
&\subjectto \; (x - x^{*})^\top F_r(x^*)\geq 0, \; \forall x: g(x)  \leq 0. \label{eqn:fGNE_NNE_set}
\end{align} 
\end{subequations}
There exist efficient solvers for variational inequalities, making the constraint~\eqref{eqn:fGNE_NNE_set} computationally tractable, even in real-time operation for a small problem size. How to ensure tractability of~\eqref{eqn:fGNE} for higher dimensional problems with multiple coupling constraints is an open research question closely linked to the theory of quasi-variational inequalities. 






\section{Case study: Electric vehicle charging game}\label{sec:CaseStudy}

Consider an electric vehicle charging game, a simplified version of~\cite{ma2016efficient, ma2013decentralized,liu2018optimal}, as played by a finite number of agents $i\in \mc{I} $. Every agent arrives with an initial charge $z^{\text{init}}_i$, and charges its vehicle by $u_i$, aiming to a desired final charge  $z^{\text{ref}}_i$. The aggregate charge, however, is constrained by $\bar U$, which is not enough to charge every agent to their desired level. This leads to the coupled optimal control problems
\begin{align}\renewcommand{\arraystretch}{1.5}
\forall i \in \mc{I} \left\{\begin{array}{rl}
\displaystyle \min_{u_i, z_i} & q_i\,\|z^{\text{ref}}_i- z_i\|^2 + (u_i)^\top \sigma_i(u)\\
 \subjectto & z_i = A_i z^{\text{init}}_i + B_i u_i,\\
 & 0\leq u_i,\quad 
  \sum_{j\in \mc{I}} u_j \leq \bar U , 
\end{array}\right.
\end{align}
where $A_i\in [0, 1]$ are the battery leakage rates, and $B_i \in [0, 1]$ is the charging efficiency. 
Similarly to~\cite{ma2013decentralized}, we model the congestion penalty $\sigma_i$ as an affine function of the total demand\footnote{A congestion penalty or price is commonly used in multi-step horizon problems to ensure peak shaving or valley filling where $\rho^1$ may be heterogeneous across agents to model different priority levels.}, i.e.,
\begin{align}
\sigma_i(u) = \Big(\rho_i^1\sum_{j\in \mc{I}} u_j + \rho_i^0\Big).
\end{align}
where $\rho_i^0$ is a base price and $\rho_i^1$ a congestion tariff. 
The parameters $q_i>0$ and $\rho_i^1, \rho_i^0>0$, respectively, indicate how much an agent values reaching their desired charge $z^{\text{ref}}_i$ relative to the incurred congestion penalty. 

%

The Lagrangian of each agent becomes:
\begin{align*}
    L_i(u,\lambda_i) = J_i(u_i, u_{-i}) + \iota_{\R_+^{n_i}}(u_i) + \lambda_i(\mathds{1}^\top u - \bar U),
\end{align*}
where $\iota_{\R_+^{n_i}}$ is the indicator function of the positive orthant, and the resulting KKT system for the set of GNEs is
\begin{align*}
\forall i\in \mc{I}:\; \left\{
\begin{array}{ll}
\nabla_{u_i}J_i(u_i, u_{-i}) + \mc{N}_{\R_+^{n_i}}(u_i) + \lambda_i=0\\
0\leq \lambda_i \perp -(\mathds{1}^\top u - \bar U)\geq 0 
\end{array}  
\right.
\end{align*}
with 
$\nabla_{u_i}J_i(u_i, u_{-i})  = 2(q_i (B_i)^2 + \rho_i^1) u_i + \rho_i^1\sum_{j\in \mc{I}\backslash \{i\}} u_j  + (-2\Delta z_i B_i + \rho_i^0)$
where 
$\Delta z_i = (z^{\text{ref}}_i - A_i z^{\text{init}}_i ) $.
Further, the set $\mc{S}^{\text{\tiny v-GNE}}$ is characterized by the following KKT system:
\begin{align*}
\begin{array}{ll}
F(u) + \mc{N}_{\R_+^{n}}(u) + \lambda \mathds{1} =  0\\
0\leq \lambda \perp -(\mathds{1}^\top u - \bar U)\geq 0 
\end{array}  
\end{align*}
with 
\begin{align*} F(u) \! = \!\!
\left[\begin{smallmatrix}
2(q_1 (B_1)^2 + \rho_1^1) & \rho_1^1 & \hdots\\
 & \ddots & \\
\rho_M^1 & \hdots &   2( q_M B_M^2 + \rho_M^1)
\end{smallmatrix}\right] \! u \! + \!\!\left[\begin{smallmatrix}
-2 \Delta z_1 B_1 + \rho_1^0\\
\vdots\\
-2 \Delta z_M B_M + \rho_M^0
\end{smallmatrix}\right]\!.
\end{align*}
Note the dependence of $F(u)$ on $\Delta z_i$ implying that the initial condition and desired final charge are measurable quantities, they are deemed comparable, and relevant for the allocation.

\subsection{Transformations}
\label{ssec:EVtransformations}

\begin{figure}[t]
\centering
\includegraphics[width= 0.95\columnwidth]{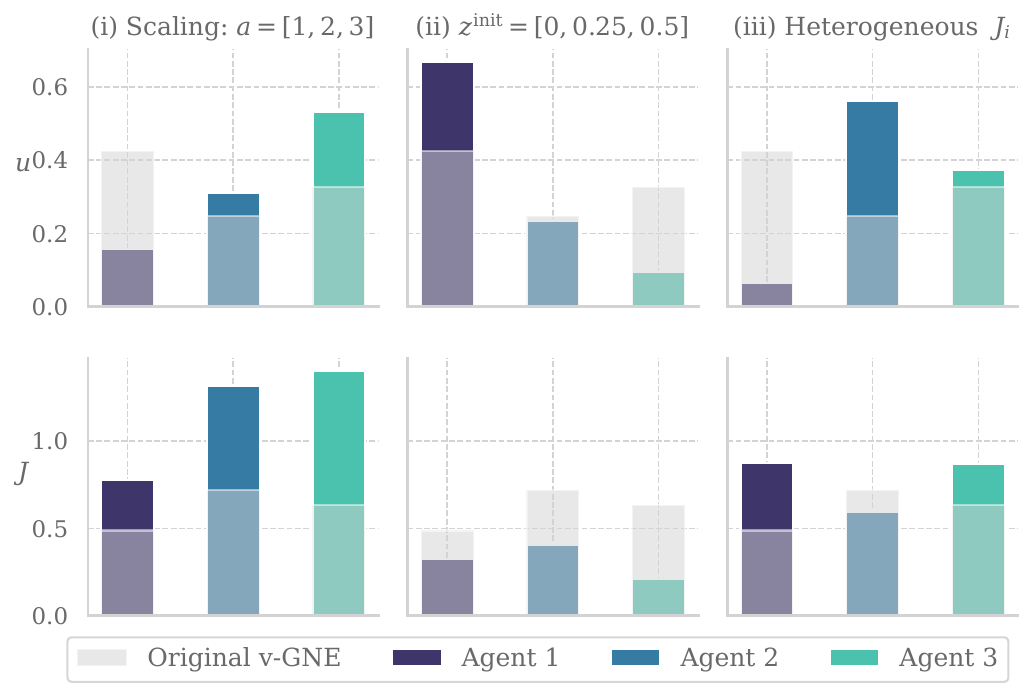}
\caption{Comparison of the v-GNE solution and how it changes under transformations of cost scaling, initial condition and cost function structure.}\label{fig:TransformationsvGNE}
\end{figure}

In Section~\ref{sec:GNEcomparability} we discussed how v-GNEs require a high level of comparability between the costs of different agents, i.e., they are sensitive to transformations of the cost functions.
As discussed in the following examples, this lack of invariance has implications for the fairness of the solution: the designer of such a system needs to consider whether the information that affects the v-GNE can be measured reliably and whether it should affect the resulting decision.
For each example, we recompute the v-GNE for the transformed costs, and we report the agents' costs and allocations in Figure~\ref{fig:TransformationsvGNE}.

\medskip

\noindent \uline{ (i) Scaling of cost functions}  $\tilde{J}_i(u) = a_i J_i(u) $ with scaling vector $a = [1\,,\,2\, ,\, 3]^\top$. Heterogeneous scaling could represent the case in which the agent's utilities depend on the additional kilometer range they get from charging their vehicle, and different vehicles have different fuel economy. Would we really deem it desirable that our solution concept allocates more charge to less efficient vehicles?

\medskip
\noindent \uline{ (ii) Change in the initial charge level} of agents from $z^{\text{init}} = [0\, , \, 0\, , \, 0]^\top$ to   $\tilde{z}^{\text{init}} = [0\, , \, 0.25\, , \, 0.5]^\top$. 
The share of charge for agent 1 who arrives with an empty battery increases significantly. However, is it really desirable that agents with lower initial charge are advantaged under the v-GNE?

\medskip

\noindent \uline{(iii) Transformed cost} - the cost function for reaching the desired charge level $z^{\text{ref}}$ is modeled as a quadratic function $J_{i} = \|z^{\text{ref}}_i- z_i\|^2$. What if the true utility is heterogeneous across agents? For instance, one agent needs to reach home but any additional charge will result in small benefit, represented by a logarithm. Another agent has to drive a long distance urgently. Consider the following cost functions:
\begin{align*} 
&J_{1}(z_1) = (\bar U-z_1)^2,\\  
&J_{2}(z_2) = \log(\frac{C}{z_2}) + 0.1 (\bar U-z_2)^2,\\ 
&J_{3}(z_3) = e^{\bar U-z_3} - 1
\end{align*}
Is it desirable that the declared need in terms of utility has such a strong effect on the allocation?

\subsection{Fairness metrics}

We now showcase the underlying comparability assumptions of a variety of fairness metrics $f(\cdot)$ and their susceptibility to changes in the scale of the cost functions. In order to do so, we compare the solution of the f-GNE in~\eqref{eqn:fGNE} under three different a-priori valid choices of fairness metrics $f(\cdot)$: (i) Nash Bargaining solution (NBS), (ii) social welfare (SW); and (iii) Maximin (MM), as introduced in Table~\ref{tab:FairnessMetrics}. 

\begin{figure}
\centering
\includegraphics[width= 1\columnwidth]{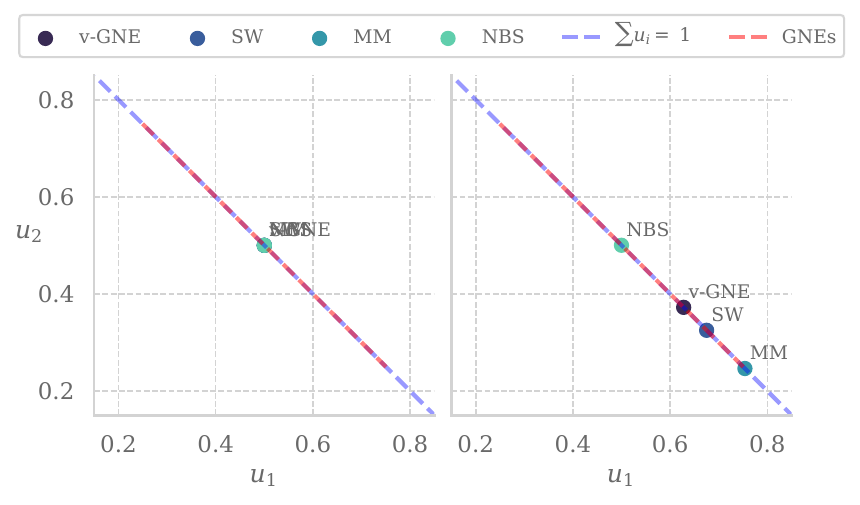}
\caption{Various fairness metrics, the v-GNE, and how their solution changes with scaling $a_1 = \{1,3\}$, left to right. Note that on the left all solutions overlap.}\label{fig:FairnessMetrics}
\end{figure}

Specifically, we consider a situation with two agents in which one unit of charge is available, i.e. $\bar U=1$, and initially both agents are equivalent and thus receive the same amount of charge $u_i = 0.5, i\in \{1,2\}$ as shown in Figure~\ref{fig:FairnessMetrics}. Then, in a second setting, we scale the cost function of the first agent by a factor of 3 such that $\tilde{J}_1(u) = a_1 J_1(u) = 3\, J_1(u)$. Based on Figure~\ref{fig:FairnessMetrics}, several observations are noteworthy:

\begin{itemize}
\item As stated in Proposition~\ref{prp:CNCofGNE}, the set of GNEs is invariant under CNC, and thus under scaling of cost functions.
\item In the scaled situation, the Maximin fairness criterion selects the GNE that allocates the most resources to agent 1 which is worse off under the scaled cost.
\item Different fairness criteria produce different allocations, with the exception of the Nash Bargaining solution, as it is invariant under CNC transformations.
\end{itemize}

\section{Discussion and Conclusion}
\label{sec:conclusions}

In a cardinal world with unit or full comparability, the v-GNE retains the desirable property of treating anonymous agents equally as they incur the same "shadow price" for accessing a resource. In such a setting, the v-GNE is interpretable as a pricing scheme that ensures economic efficiency,
akin to how a single clearing price emerges from bids and asks in a competitive market.



However, in worlds where the cost functions of agents are less than cardinal unit comparable, 
one needs to be cautious when selecting the v-GNE with fairness goals in mind. 

This will happen, for example, whenever an agent self-declares their cost, which comes with an incentive to misrepresent them for private benefit. 
%
%
Concerns with the truthfulness and accuracy of reporting are not the only problems, as certain comparability levels require access to relevant information regarding the agents' costs. In practice, such information may be hard to quantify, expensive to measure, or agents may not want to share such information for privacy concerns. Sometimes, the designer may wish to deliberately exclude some features from influencing the allocation and make solutions invariant to some cost transformations.

In sum, the fair nature of v-GNEs is conditioned on a high level of interpersonal comparability that the designer needs to verify, and is related to a precise notion of fairness. 
Other fairness metrics require less comparability; in particular, the Nash bargaining solution is attractive for this reason. The general problem of selecting a GNE based on a fairness metric can be formulated as a bilevel optimization which is computationally tractable for problems with one coupling constraint. Higher dimensional cases require further study.




\bibliography{GNE_comparability_CDC2025}

\clearpage
\appendix
\subsection{Proposition~\ref{prp:CNCofGNE}}\label{sec:App_Prop1}

Using the transformed cost of all agents $ \tilde{J}_i(x) = \phi_{\text{\tiny CNC}} (J_i(x)) = a_i J_i(x) + b_i$ with $a_i>0$ and $b_i \in \R$, we define an auxiliary GNE problem as follows
\begin{equation}\label{eq:GNEPauxiliary}
\forall i\in \mc{I}: \;  \min_{x_i}\, {\tilde{J}_i(x_i, x_{-i})} \subjectto x_i \in \mc{C}_i(x_{-i}),
\end{equation}
and we denote the corresponding solution set as $\tilde{\mc{S}}^{\text{\tiny GNE}}(x)$.

Assume that $x^{*}\in \mc{S}^{\text{\tiny GNE}}$ is a solution of the original GNE~\eqref{eq:GNEP1}, thus by~\cite[Theorem 4.6 (a)]{facchinei2007gnep} there exists 
$\{\lambda_i\}_{i \in \mc{I}}$ 
fulfilling the KKT conditions:
\begin{align}\label{eq:KKToriginalGNE}
\forall i\in \mc{I}:\; \left\{
\begin{array}{ll}
\nabla_{x_i}J_i(x^{*}) + \nabla_{x_i}g_i(x^*)^\top \lambda_i^*=0\\
0\leq \lambda_i^* \perp -g_i(x^*)\geq 0 .
\end{array}
\right. 
\end{align}
We multiply the stationary condition in~\eqref{eq:KKToriginalGNE} by $a_i> 0$
\begin{align*}
a_i \nabla_{x_i}J_i(x^{*}) +\nabla_{x_i} g_i(x^*)^\top(a_i \lambda_i^*)=0, 
\end{align*}
and define $\tilde{\lambda}_i^* = a_i \lambda_i^*$. Each pair $(x_i^*, \tilde{\lambda}_i^*$) fulfills the KKT conditions
\begin{align}\label{eq:KKTtransformedGNE}
\forall i\in \mc{I}:\; \left\{
\begin{array}{ll}
 \nabla_{x_i}\tilde{J}_i(x^{*}) + \nabla_{x_i}g_i(x^*)^\top\tilde{\lambda}_i^*=0\\
 0\leq \tilde{\lambda}_i^* \perp -g_i(x^*)\geq 0. 
 \end{array}
\right. 
\end{align}
At $x^*$ an appropriate constraint qualification holds by assumption for all agent's subproblems in~\eqref{eq:KKToriginalGNE}, thus it also holds at $x^*$ for~\eqref{eq:KKTtransformedGNE}. Consequently, by~\cite[Theorem 4.6 (b)]{facchinei2007gnep} any $x^*\in \mc{S}^{\text{\tiny GNE}}$ is also a GNE of the auxiliary GNE problem~\eqref{eq:GNEPauxiliary}. The inclusion in the other direction is proved similarly, and thus $\tilde{\mc{S}}^{\text{\tiny GNE}} = \mc{S}^{\text{\tiny GNE}}$.

\subsection{Proof of Proposition \ref{prp:CUCofVGNE}}\label{sec:App_Prop2}

The first statement of Proposition~\ref{prp:CNCofGNE} (that $\mc{S}^{\text{\tiny v-GNE}}$ is not invariant under $\phi_{\text{\tiny CNC}} (\cdot)$ can be proven by repeating the first steps in the proof of Proposition~~\ref{prp:CNCofGNE}. To prove the second statement in the proposition we define the transformed cost as $\tilde{J}_i(x) = \phi_{\text{\tiny CUC}} (J_i(x)) = a J_i(x) + b_i$ with $a>0$ equal for all agents. Let $x^{*}\in \mc{S}^{\text{\tiny v-GNE}}$ , then the KKT conditions associated with the GNE problem in~\eqref{eq:GNEP1} hold with the multipliers and the constraint function being equal for all agents~$\forall i\in \mc{I}: \lambda_i^* = \lambda^{*}\in \R^{m}_{+}, \,$ and $g_i = g:  \R^n \to \R^m$ are given by:
\begin{align}\label{eq:KKToriginalvGNE2}
\forall i\in \mc{I}:\; \left\{
\begin{array}{ll}
\nabla_{x_i}J_i(x^{*}) + \nabla_{x_i}g(x^*)^\top \lambda^{*}=0\\
0\leq \lambda^* \perp -g(x^*)\geq 0 
\end{array}
\right. 
\end{align}
we multiply the stationary condition in~\eqref{eq:KKToriginalvGNE2} by $a> 0$
\begin{align*}
a \nabla_{x_i}J_i(x^{*}) +\nabla_{x_i} g(x^*)^\top (a\lambda^*)=0, 
\end{align*}
and define $\tilde{\lambda}^{*} = a \lambda^{*}$. Each pair $(x_i^*, \tilde{\lambda}^{*}$) fulfills
\begin{align}\label{eq:KKTtransformedvGNE}
\forall i\in \mc{I}:\; \left\{
\begin{array}{ll}
 \nabla_{x_i}\tilde{J}_i(x^{*}) + \nabla_{x_i}g(x^*)^\top (\tilde{\lambda}^{*})=0\\
 0\leq \tilde{\lambda}^* \perp -g(x^*)\geq 0. 
 \end{array}
\right.
\end{align}
which is a sufficient condition for $x^* \in \mc{S}^{\text{ \tiny v-GNE}}$ which thus remains invariant under transformations of the class $\phi_{\text{\tiny CUC}}(\cdot)$. The inclusion in other direction is proved similarly.

\subsection{Proof of Proposition~\ref{prp:vGNEequalfGNE}}\label{sec:App_Prop3}

Consider the optimization problem  
\begin{align}\label{eq:OPfair}
x^{\text{\tiny OP}} = \argmin_{x} f(x) \quad \subjectto g(x)\leq 0 
\end{align}
with $f(x)  = \sum_{i\in \mc{I}} J_i(x_i)$ and the KKT conditions
\begin{align*}
\left\{
\begin{array}{ll}
\nabla_x f(x^{\text{\tiny OP}} ) + \nabla_x g(x^{\text{\tiny OP}} )^\top  \lambda  \\
= \vcol(\{\nabla_{x_i}J_i(x^{\text{\tiny OP}} )\}_{i\in \mc{I}}) + \vcol(\{  \nabla_{x_i} g(x^{\text{\tiny OP}} )^\top  \lambda\}_{i\in \mc{I}}) =0\\[5pt]
0\leq \lambda^* \perp -g(x^{\text{\tiny OP}} )\geq 0. 
\end{array}\right.
\end{align*}
which coincides with the KKT conditions of the v-GNE solution~\eqref{eq:KKToriginalvGNE} for $J_i(x_i)$. Thus, we have that $x^{\text{\tiny OP}} = x^{\text{\tiny v-GNE}}$ and we know that $x^{\text{\tiny v-GNE}} \in \mc{S}^{\text{\tiny GNE}}$. As a consequence, we can state~\eqref{eqn:fGNE}
\begin{align*}
x^{\text{\tiny f-GNE}} = \argmin_{x^*}& \; f(x^*) \subjectto  x^* \in  \mc{S}^{\text{\tiny GNE}}, \quad g(x^*)\leq 0 
\end{align*}
which is the optimization problem in~\eqref{eq:OPfair} with the additional GNE constraint. However, we have previously established that the optimal value $x^{\text{\tiny OP}}$ of ~\eqref{eq:OPfair} is attained at the v-GNE and thus $x^{\text{\tiny v-GNE}} =  x^{\text{\tiny f-GNE}}$.

\end{document}